\documentstyle[psfig,preprint,aps]{revtex}

\catcode`\@=11

\def\big#1{{\hbox{$\left#1\vbox to11.5\p@{}\right.\n@space$}}}
\def\Big#1{{\hbox{$\left#1\vbox to14.5\p@{}\right.\n@space$}}}
\def\bigg#1{{\hbox{$\left#1\vbox to17.5\p@{}\right.\n@space$}}}
\def\Bigg#1{{\hbox{$\left#1\vbox to21.5\p@{}\right.\n@space$}}}
\def\BIG#1{{\hbox{$\left#1\vbox to25.5\p@{}\right.\n@space$}}}

\catcode`\@=12

\newcommand{\mathchoices}[2]{{\mathchoice{#1}{#1}{#2}{#2}}}

\newcommand{\lab}[1]{\protect\label{#1}}
\newcommand{\refb}[1]{(\protect\ref{#1})}
\newcommand{\refeq}[1]{eq.~(\protect\ref{#1})}

\newcommand{\bold}[1]{\mbox{\protect\boldmath $#1$}}
\newcommand{\vc}[1]{\mathchoice{\bold{#1}}{\bold{#1}}{\vec{#1}}{\vec{#1}}}

\newcommand{\be}{\begin{equation}}
\newcommand{\ee}{\end{equation}}
\newcommand{\bea}{\begin{eqnarray*}}
\newcommand{\eea}{\end{eqnarray*}}
\newcommand{\beagrp}[1]{\begin{equation}\protect\lab{#1}\begin{minipage}{14cm}
      \topsep0mm\abovedisplayskip0mm
      \belowdisplayskip0mm\begin{eqnarray*}}
\newcommand{\eeagrp}{\end{eqnarray*}\end{minipage}\end{equation}}
\newcommand{\beano}{\begin{eqnarray}}
\newcommand{\eeano}{\end{eqnarray}}

\newcommand{\ibox}[1]{{\mbox{\rm\scriptsize #1}}}
\newcommand{\hart}{{}^\ibox{hard}}
\newcommand{\weich}{{}^\ibox{soft}}
\newcommand{\ret}{{}^\ibox{ret}}

\newcommand{\RR}{\mathchoices{{\sf I\hskip-.10em R}}{
			     {\mbox{\footnotesize\sf I\hskip-.10em R}}}} 

\newcommand{\GeV}{\,Ge\!\hskip0.1ex V}
\newcommand{\fm}{\,f\!\hskip0.1ex m}

\newcommand{\cM}{{\cal M}}

\newcommand{\hint}[2]{\underbrace{#1}_{\makebox[0pt]{\scriptsize #2}}}

\newcommand{\Ei}{\mathop{\rm Ei}\nolimits}

\renewcommand{\Im}{\mathop{\sl Im}\nolimits\;}

\begin{document}
\bibliographystyle{dstt}

\title{Hard Photon Production Rate of a Quark-Gluon Plasma at Finite Quark
Chemical Potential\footnote{supported by BMFT and GSI Darmstadt}}
\author{Christoph T. Traxler, Hans Vija\footnote{present address:
Physics Department, University of Washington, Seattle, WA 98195, USA},
and Markus H. Thoma}
\address{Institut f\"ur Theoretische Physik, Universit\"at Giessen,\\
35392 Giessen, Germany\footnote{e-mail: Chris.Traxler@uni-giessen.de,
vija@u.washington.edu, thoma@theorie.physik.uni-giessen.de}}
\maketitle

\begin{abstract}
We compute the photon production rate of a quark-gluon plasma (QGP) at finite
quark chemical
potential $\mu $ using the Braaten-Pisarski method, thus continuing the
work of Kapusta, Lichard, and
Seibert who did the calculation for $\mu=0$.
\end{abstract}

The thermal production of hard photons in ultrarelativistic heavy ion
collisions has been
proposed as a possible signature for the formation of a quark-gluon plasma
(QGP) \cite{ruu92}.
Since the mean free path of photons in the fireball is much larger than its
dimensions \cite{tho94}, photons
provide a direct probe of the fireball. So far the production
rate of hard photons
has been considered at finite temperature but vanishing quark chemical
potential mainly. In
order to include medium effects consistently the Braaten-Pisarski resummation
technique \cite{bp90a} has been applied to this problem \cite{kls91,bai92}.

Assuming the formation of a QGP already at AGS and SPS energies, however,
a finite quark chemical potential has to be considered \cite{nag92}, and even
at RHIC energies the quark chemical potential $\mu$ may not be
negligible as
indicated by RQMD simulation ($\mu\approx1-2\> T$) \cite{dsr92}. At given
energy
density estimates
based on lowest order perturbation theory indicate a strong suppression of
the photon
production at non-vanishing $\mu $ as compared to the case $\mu =0$
\cite{dum93}. The aim of the present work is the
improvement of these
estimates by applying the Braaten-Pisarski method generalized to finite
chemical potential \cite{vij94}.

We follow the calculation of Kapusta, Lichard, and Seibert \cite{kls91}, and
describe the changes that have to be made for a nonzero quark
chemical potential. For
brevity, we do not demonstrate the intermediate steps of the computation
where
they would be analogous to ref. \cite{kls91}.

To leading order the hard photon production rate is derived by using
the Braaten-Yuan prescription \cite{by91} resulting in a decomposition
into a soft part, which is treated using the resummed propagators of
Braaten and Pisarski, and a hard part containing
only bare propagators. For this purpose a parameter $k_c$ is
introduced, separating the soft from the hard momenta of the
intermediate quark. Demanding $gT\ll k_c\ll T$ and assuming
the weak coupling limit, the final result is independent of the separation
scale $k_c$.

The soft part can be obtained from the imaginary part of the
self energy of a photon propagating through a QGP. The contributing diagrams
to the photon self energy are shown in fig.~1.
There quark lines with blobs represent effective quark
propagators \cite{bpy90},
however for the $\mu\neq0$--case with a modified quark mass \cite{vij94,bp92}
of
\be
m_q^2={g^2\over6}\left(T^2+{\mu^2\over\pi^2}\right)\quad.
\ee
Our result for the soft contribution to the production rate of a hard photon
with energy $E$ and momentum ${\vc p}$ is
\beano
  2E{dR\weich\over d^3p}
   &=&-{2\over(2\pi)^3}g^{\mu\nu}\Im\Pi\ret_{\mu\nu}{1\over e^{E/T}-1}
   \nonumber\\
   &=&{5\alpha\alpha_s e^{-E/T}\over9\pi^2}\left(T^2+{\mu^2\over\pi^2}\right)
     \ln\left({k_c^2\over 2m_q^2}\right)\quad.\lab{sp}
\eeano
This differs from the result obtained in ref. \cite{kls91} only
in that (a) the quark
mass $m_q$ depends now on the chemical potential and (b) the factor in front
of the logarithm is modified since it
originates from a factor $m_q^2$. (In deriving \refeq{sp} we assumed a
hyperbolic cut-off $k_c^2>\vc k^2-\omega ^2>0$ in accordance to the
hard part \refb{hp} following ref. \cite{kls91}, where $\omega $ is the
energy and $\vc k$ the momentum of the spacelike intermediate soft quark.
Thus $\omega$ and $|\vc k|$ need not to be soft individually. However,
closer investigations show that for the expression under the loop integral
in $\Pi _{\mu \nu}^{ret}$, approximations are legitimate that
assume $\omega$ as well as $|\vc k|$ to be soft, since it is this region
where the main contribution to the loop integral comes from.)

The easiest way to calculate the hard part is starting from the scattering
matrix elements shown in fig.~2. They are related to the photon self energy
of fig.~1 by cutting the latter \cite{wel83}.
{}From QCD Feynman rules, we find the squared matrix elements to be
\be
\Sigma{|\cM|^2}={2^9\cdot5\over9}\pi^2\alpha\alpha_s{u^2+t^2\over ut}
\ee
for the annihilation process, and
\be
\Sigma{|\cM|^2}=-{2^9\cdot5\over9}\pi^2\alpha\alpha_s{s^2+t^2\over st}
\ee
for each of the two Compton processes for quarks and antiquarks. The
symbol  $\Sigma$ indicates
that these matrix elements are already summed over spins, colors, and two
flavors ($u$ and
$d$). The letters $s$, $t$, and $u$ denote the Mandelstam variables.
In each diagram, all legs but the outgoing photon belong to thermalized
particles. The photon
is not thermalized since it has a large mean free path. For each of the
three possible processes we may compute the hard part of the photon
production rate as \cite{kls91}
\be
2E{dR\hart\over d^3p}={1\over(2\pi)^8}\int{d^3p_1\over2E_1}
{d^3p_2\over2E_2}{d^3p_3\over2E_3}
\delta^4(P_1+P_2-P_3-K)n_1(E_1)n_2(E_2)(1\pm n_3(E_3))\Sigma{|\cM|^2}
\ee
where $n_{1,2,3}$ are Bose or Fermi distribution functions, respectively;
the plus sign is
for the annihilation process and the minus for the two Compton processes.
Still following ref. \cite{kls91}, we may rewrite this equation as
\be
\lab{hp} 2E{dR\hart\over d^3p}={1\over8(2\pi)^7E}
  \int\limits_{2k_c^2}^\infty ds\int\limits_{-s+k_c^2}^{-k_c^2} dt\,
  \Sigma{|\cM|^2}
  \int\limits_{\RR^2}dE_1\,dE_2 {\Theta(P(E_1,E_2))n_1n_2(1\pm n_3)
  \over\sqrt{P(E_1,E_2)}}
\ee
with the polynomial $P(E_1,E_2)=-(tE_1+(s+t)E_2)^2+2Es((s+t)E_2-tE_1)
-s^2E^2+s^2t+st^2$ and the step function $\Theta $.

For the $\mu=0$--case, it is a reasonable approximation
to use Boltzmann distribution functions for $n_1$ and $n_2$ instead of the
full Fermi/Bose
functions. In this way, all integrations may be performed analytically,
and the result
\be
2E{dR\over d^3p}={5\over9}{\alpha\alpha_s T^2e^{-E/T}\over\pi^2}
   \bigg(\hint{{2\over3}\ln\left({4ET\over k_c^2}\right)-1.43}
   {annihilation}
   +\hint{{1\over3}\ln\left({4ET\over k_c^2}\right)+0.015}{Compton}\bigg)
\ee
fits perfectly together with the soft part: the sum of both is independent
of $k_c$.

Actually, the Boltzmann approximation (for $n_1$ and $n_2$) is better than it
should
be, considered that the
photon contribution from the Compton processes is significantly underestimated
(up to 30\%)
and the annihilation contribution is overestimated. Surprisingly, both errors
cancel
each other, up to an error of about 10\% in the final result for those values
of $T$, $E$, and $g$ that are interesting in practice.

The total photon production rate for $\mu=0$, computed in the Boltzmann
approximation (for $n_1$, $n_2$), is
\be
\lab{rate0}
   2E{dR\over d^3p}={5\over9}{T^2\alpha\alpha_s e^{-E/T}\over\pi^2}
   \ln\left(2.91E\over g^2T\right) \quad,
\ee
which also follows from the photon damping rate by the principle of
detailed balance \cite{tho94}.

Dumitru et al. \cite{dum93} use the Boltzmann approximation also in their
computation
of the photon rate at $\mu\neq0$, however they obtain a hard part which
does not match onto the
analytically known soft part, (\ref{sp}). The cited result also contains a
term
$\sim\Ei\left((4E\mu-k_c^2)/4ET\right)$ which runs over a branch point at
finite $\mu$.

Our approach to the problem is the following: since we cannot evaluate the
integrals in (\ref{hp}) containing the exact distribution functions
analytically, we employ numerical methods. We rewrite (\ref{hp}) in a form
suitable for Gauss quadrature ($E_+:=E_1+E_2$):
\beagrp{numform}
  2E{dR\hart\over d^3p}
  &=&-{5\alpha\alpha_s\over18\pi^5E}e^{-E/T-k_c^2/2ET}
  \ \hint{\int\limits_{2k_c^2}^\infty ds\,e^{-(s-2k_c^2)/4ET}}{Laguerre}
  \quad{1\over s}
  \hint{\int\limits_{-s+k_c^2}^{-k_c^2} dt}{Legendre}\,|\cM(s,t)|^2\times\\
  &&\times\hint{\int\limits_{E+s/4E}^\infty dE_+\,e^{-(E_+-E-s/4E)/T}}
  {Laguerre}
  {1\over 1\mp e^{-(E_+-E-\mu_3)/T}}\times\\
  &&\times\hint{\int\limits_{E_2^-}^{E_2^+}{dE_2\over\sqrt{P_1(E_+,E_2)}}}
  {Chebyshev}
  {1\over e^{-\mu_1/T}\pm e^{-(E_+-E_2)/T}}{1\over e^{-\mu_2/T}+e^{-E_2/T}}
  \quad.
\eeagrp
In this compact notation, the upper signs and $\mu_1=-\mu_2=\mu$, $\mu_3=0$
apply to the
annihilation process; the Compton processes require the lower signs and
$\mu_1=0$,
$\mu_2=\mu_3=\pm\mu$, where the two results for $+\mu$ and $-\mu$ have
to be added in order to
take antiquarks as well as quarks into consideration.
The polynomial $P_1(E_+,E_2)$ is just $P(E_+-E_2,E_2)/s^2$;
thus it has the leading coefficient
$-1$ and gives under the square root a perfect weight for a Gauss-Chebyshev
quadrature
in $E_2$. The $E_+$--integral, as well as the $s$--integral, is done
numerically by a
Gauss-Laguerre quadrature; the necessary exponential weight function arises
naturally in the
expression. The $t$--integral, having no appropriate weight function, is
performed via
Gauss-Legendre quadrature. Finally, the $s$--integral is done by a
Gauss-Laguerre quadrature, as suggested again by the naturally arising weight
function.

Unfortunately, each of the four integrands is highly peaked at the ends of
the integration
interval, due to singularities in or near the domain of integration. This
means
the integral value is dominated by contributions coming from comparably small
regions, and the
quadrature problem is ill-conditioned.
In order to cure the problem, we subtract from the integrand in each step of
the fourfold integration those contributions that stem from poles inside or
slightly outside
the integration region. Those parts are integrated out separately in an
analytical way; the
remaining numerical integrals are much easier to compute, since the
integrands are no longer
varying strongly. This way, we need only about 20 points for each quadrature,
and still obtain numerical results for the hard part that are precise up to
an maximal error of $0.2\%$.

In fig.~3 $k_c$ is varied over almost six orders of
magnitude, leaving the total photon rate fixed within numerical error.
(In order to test the cancellation of $k_c$, expected to occur in the
weak coupling limit, we adopt a value of $g=0.01$.)
Even in regions where $gT\ll k_c\ll T$ is by no means true, the sum of
soft and hard part
is perfectly constant, while the soft part alone varies over a wide range
and even becomes negative.
This is a nice verification of the Braaten-Yuan method for $\mu\neq0$
\cite{vij94,by91}.
{}From (\ref{sp}) and the independence on $k_c$ we know that the final result
has to assume the form
\be\label{ff}
2E{dR\over d^3p}\ =\ {5\alpha\alpha_s e^{-E/T}\over9\pi^2}
   \left(T^2+{\mu^2\over\pi^2}\right) \left(\ln{2.91ET\over
   g^2(T^2+\mu^2/\pi^2)}+G\right)\quad ,
\ee
where the dimensionless quantity $G$ follows from the hard part.
As a result of the numerical calculation of the hard part, discussed
above, it turns out that $G$ depends only on $\mu/T$.
Within a $3\%$ error in the rate, $G$ depends
very weakly on $E/T$. However, $G$ is nicely independent of $T$, as a
dimensionless quantity should be. We found that $G$ can be fitted to a
good approximation by the phenomenological
formula $G=\ln(1+\mu^2/\pi^2T^2)$, thus leading to the compact expression
\be
\lab{phen}
 2E{dR\over d^3p}\ =\ {5\alpha\alpha_s e^{-E/T}\over9\pi^2}
   \left(T^2+{\mu^2\over\pi^2}\right) \ln\left({2.91E\over g^2T}\right)\quad.
\ee
This pocket formula reproduces the exact, numerically calculated result
within an error
of $3\%$ for $|\mu/T|\le 1$. A larger $\mu/T$ requires a one-parameter fit for
$G$. We found that $G=\ln(1+0.139\mu^2/T^2)$, inserted in \refeq{ff}, leads
to a phenomenological formula for the rate that is precise on the $3\%$-level
even for much larger $|\mu/T|$. We emphasize that the error of $3\%$ is not due
to the numerical evaluation of the hard part but to the assumption that $G$ is
solely dependent on $\mu/T$.

We are now able to extrapolate this formula to realistic values of $g$ and
discuss the photon spectrum of the QGP. For photon energies $E$ above about
$3T$ the logarithm in (\ref{phen}) is positive, indicating the validity
of the extrapolation to realistic values of the coupling constant
\cite{tho93}.
Fig.4 shows how the spectra are dominated by the exponential decrease with
temperature. We draw the conclusion that a measurement of the photon spectrum
may serve only or
mainly to find out the temperature of the plasma; in order to measure the
chemical potential by
a photon experiment, one would have to measure the overall coefficient of
the exponential, which
is in an experimentally obtained spectrum influenced by many other factors
not yet under control, like the size and duration of the plasma phase.
However, if one has information on the energy density $\epsilon$ of the
plasma, $T$ and $\mu$ are related to each other
by an equation of state like the one quoted
in ref. \cite{dum93}:
\be
\epsilon=\left({37\pi^2\over30}-{11\pi\alpha_s\over3}\right)T^4
 +3\left(1-{2\alpha_s\over\pi}\right)T^2\mu^2
 +{3\over2\pi^2}\left(1-{2\alpha_s\over\pi}\right)\mu^4+(0.2\GeV)^4\quad.
\ee
If $T$ is made dependent on $\mu$ in this fashion \cite{dum93}, the resulting
photon spectra are
strongly dependent on $\mu$. At RHIC, one expects a maximum energy density
of about
$\epsilon=5\GeV/\fm^3$. In fig.~5, we show the corresponding photon spectra.
The temperatures of the five curves decrease with rising $\mu$ from
$0.27\GeV$ to about
$0.22\GeV$, whereas $\mu/T$ varies from $0$ to about $1.8$.
The photon suppression at finite chemical potential and fixed energy density
observed in ref. \cite{dum93} is therefore an
indirect phenomenon, caused by the reduction of the temperature.

In conclusion, at $\mu=0$, our result (\ref{phen}) conincides with the one in
ref. \cite{kls91}. The Boltzmann approximation used in there seems to work
fine.

At $\mu\neq0$, Bose and Fermi distributions have to be used in order to match
the soft contribution onto the hard one according to the Braaten-Yuan
prescription providing a consistent result to leading order. Hence the hard
part of the photon production rate can only be determined numerically.
We were able to demonstrate that the final result is independent of the
arbitrary separation parameter
$k_c$ and obtained the simple formula (\ref{phen}).
Our graphs of the photon spectrum have a strong
similarity to the one of Dumitru et al. \cite{dum93}, mainly because they all
are dominated by the single factor $e^{-E/T}$.

We did not consider pre-equilibrium effects so far. The equilibrium
distribution functions used in our
calculation are not quite applicable, since the plasma is probably never
completely equilibrated.
For a phenomenological treatment of a pre-equilibrium plasma, fugacity
factors may be employed
\cite{bir93}. This will be dealt with in a future publication.

\acknowledgements
We would like to thank R.~Baier, T.~S.~Bir\'o, A.~Dumitru, P.~Lichard,
D.~Rischke, and D.~Seibert for useful discussions.


\end{document}